# Investigation of the Impact of Magnetic Fields on Scattering Muography Images

Hamid Basiri,[1] Tadahiro Kin,[1] Eduardo Cortina Gil,[2] and Andrea Giammanco[2]

[1]*Interdisciplinary Graduate School of Engineering Science, Kyushu University, 6-1 Kasuga-koen, Kasuga-shi, Fukuoka 816-8580, Japan*

[2]*Centre for Cosmology, Particle Physics and Phenomenology, Université catholique de Louvain, Chemin du Cyclotron 2, B-1348 Louvain-la-Neuve, Belgium*

*Corresponding author: Hamid Basiri*
*Email: basiri.hamid@gmail.com*

## Abstract

Muography is a noninvasive imaging technique that exploits cosmic-ray muons to probe various targets by analyzing the absorption or scattering of muons. The method is particularly useful for applications ranging from geophysical exploration to security screening, including the identification of nuclear materials. This study leverages both Monte Carlo simulations and the Point of Closest Approach (PoCA) algorithm for image reconstruction to specifically explore the distortions caused by magnetic fields in scattering muography images. In the PoCA algorithm, it is assumed that all scattering of a muon during its travel in material occurs at a single point, known as the PoCA point. Each PoCA point is characterized by a scattering angle, whose distribution provides insights into the density and elemental composition of the target material. However, magnetic fields can influence muon trajectories according to Lorentz's law, affecting the estimated positions of the PoCA points and the calculated scattering angles. This introduces challenges in applications such as border security control systems. Moreover, the presence of magnetic fields can lead to what we term "magnetic jamming", where the resulting muography image is distorted or misleading. This effect further complicates the accurate identification and interpretation of target materials. Our findings underline the necessity to account for magnetic field distortions when utilizing scattering muography in practical scenarios.

## 1. INTRODUCTION

The discovery of cosmic rays by Victor Hess over a century ago [1], followed by the identification of muons by Anderson and Neddermeyer [2], marked a significant milestone in the field of particle physics. Cosmic-ray muons continuously bombard the Earth's surface at a rate of roughly 10,000 per square meter per minute and have been deployed for a variety of applications [3]. These applications benefit from the unique properties of these particles, including the ability to penetrate deeply into any material, their universal availability (even underground), very high detection efficiency, and nondestructive interaction with objects. Such properties make them invaluable in applications to both fundamental (e.g., LHC experiments) and applied research. One notable technique that leverages these properties is muon radiography, also known as muography [4]. This technique enables the visualization of the internal structure of various targets through two main methods: by measuring the absorption ratio, which gives information on the density profile of the object, and by measuring the average Coulomb scattering angle, which provides insight into the atomic number of the material under investigation (with applications mostly in the nuclear sector, where one is interested in identifying special nuclear materials whose atomic number is very large). Muon radiography was first employed by Alvarez to study hidden chambers within pyramids [5]. Since then, muon imaging applications have expanded to fields such as volcanology [6], geotomography [7], nuclear waste investigations [8], inspection of nuclear reactors [9], detection of shielded high-Z materials [10], imaging of cultural heritage objects [11], infrastructure degradation investigation [12], and 3D muon tomography of targets using portable detectors [13].

Recent work by our group has delved into the use of cosmic-ray muons for magnetic field investigation [14, 15, 16, 17, 18, 19]. In these studies, muons are deflected in the magnetic field based on factors such as their energy, the density of the magnetic field flux, and the extension of the magnetized volume. These deflections are governed by the Lorentz force. Preliminary results have shown that muography offers promising solutions to some of the limitations we encounter with existing techniques, especially when dealing with large and strong magnetic fields.

In this study, we focus on examining the impact of magnetic fields on scattering muography, a technique where the scattering angle distribution is crucial for determining the material's density and atomic number ($Z$). This dependency is especially significant for real-world applications such as nuclear material identification and border security [20]. However, magnetic fields introduce complications by altering muon trajectories through the Lorentz force, an aspect that remains underexplored. By investigating these magnetic field-induced distortions, our study aims to enhance the utility of scattering muography, particularly in environments where magnetic fields are present.

## 2. MATERIALS AND METHODS

We employ the Particle and Heavy Ion Transport code System (PHITS) for our simulations [21]. The design for the high-Z material identification system is based on the studies by P. La Rocca et al. in the Muon Portal Project [22], which is representative of the most approaches in border security applications of muography [20]. For generating cosmic-ray muons, we utilized the PARMA model [23], integrated within the PHITS Monte Carlo code. We generated muons according to the sea-level flux expected at the latitude and longitude of the Chikushi Campus, Kyushu University, as of January 1st, 2020. This simulation mirrors a real-world measurement span of 13 minutes. Although it is true that the longer the acquisition time is, the better the image can be interpreted, in cargo scanning and other security applications a trade-off must be found with the maximization of the number of scanned cargos (or other objects) per day. We have chosen the measurement time based on the Muon Portal Project [22] and aimed to simulate a realistic scanning scenario, which is ideally less than 10 minutes. The selection of 13 minutes allows for a slight increase to explicitly emphasize the effects of magnetic fields on muon trajectories, enhancing the visibility of these effects within the constraints of a realistic scan duration.

For muon trajectory reconstruction, we utilized the Point of Closest Approach (PoCA) algorithm [24]. This algorithm determines the shortest distance between the muon's entry and exit within the observation region. Magnetic fields can shift the PoCA point and the scattering angle, which is crucial for the interpretation of muography in magnetically influenced areas. The PoCA principle and the magnetic field's influence on it are illustrated in Figure 1.

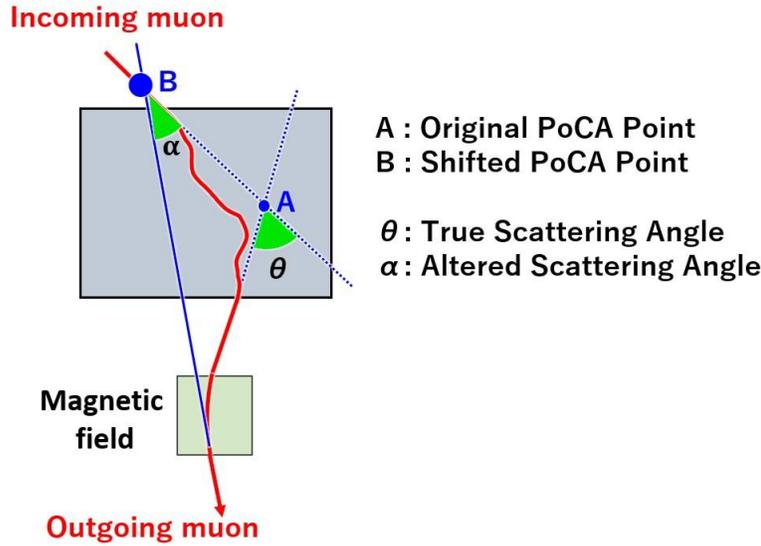

FIGURE 1: Schematic depiction of the PoCA algorithm and of the distortion due to magnetic field influence.

In the PoCA method, muon scattering events are assumed to occur at a single point, termed the PoCA point. For each muon, the distance between its incoming and outgoing trajectories is calculated. The midpoint of this distance represents the PoCA point, where scattering is assumed to have taken place. Additionally, by evaluating the directions of incoming and outgoing muons, the scattering angle is determined, providing insights into the material's density or atomic number ($Z$).

To study the effects of a magnetic field on scattering muography, we modeled a target shaped like the word "MUON" in our PHITS simulations. This target, 40 cm thick and centrally placed in a standard 20-foot shipping container (dimensions 600 cm × 234 cm × 228 cm), consists of letters made from materials with varying atomic numbers: uranium for "M", lead for "U", iron for "O", and aluminum for "N". These materials were chosen to cover a broad range of atomic numbers. The detection system comprises two layers of plastic scintillator, each 3 m by 6 m, positioned both below and above the detection area, as illustrated in Figure 2. For the purposes of this simulation, we assume the detectors have perfect resolution. This assumption allows us to isolate and focus on the effects of the magnetic field, eliminating the complicating factors of detector imperfections such as finite resolution, efficiency variances, and background noise. This idealized detector setup ensures that our study concentrates on the intrinsic changes in muon trajectories and scattering angles induced by magnetic fields.

To accurately study the effects of a magnetic field on muon scattering, the "MUON" target, designed to be 40 cm thick, is centrally placed within a magnetic field extending to 50 cm in thickness. This configuration ensures that there is an additional 5 cm of magnetic field coverage both above and below the target. The choice of a 40 cm target thickness was motivated by the need to clearly create and resolve the shape of the letters "M", "U", "O", and "N" within the simulation, allowing for a clear distinction in the imagery in the absence of a magnetic field. The extra 5 cm of magnetic field coverage on each side of the target ensures uniform magnetic influence throughout the target volume, which is crucial for the consistency and accuracy of our measurements.

The length of the magnetic field, precisely aligned to cover the target's position within the container, is essential for ensuring that all muons traversing the target are uniformly influenced by the magnetic field. This uniformity is crucial for analyzing how the magnetic field affects muon paths and scattering angles. The additional coverage beyond the target dimensions minimizes edge effects and guarantees that the muon paths are fully influenced by the magnetic field, enabling a direct assessment of how magnetic fields impact muon scattering properties.

The precise positioning of the magnetic field, with a careful 5 cm extension beyond the target's top and bottom edges, is meticulously chosen to align with the target area. This alignment ensures that muons passing through the target are consistently subjected to the magnetic field's influence, allowing for a controlled investigation into the magnetic field's effects on muon behavior and scattering patterns.

First, we obtained an image of the "MUON" target using the PoCA method. Then, the target was enclosed within a uniform dipole magnetic field. We progressively increased the magnetic field's strength from 100 mT to 500 mT, in 100 mT increments, to observe how different magnetic field strengths affect the PoCA points and angles, thereby influencing material identification accuracy. Images of the target were captured at each increment of magnetic field strength, facilitating a comprehensive analysis of the magnetic field's impact on muon scattering.

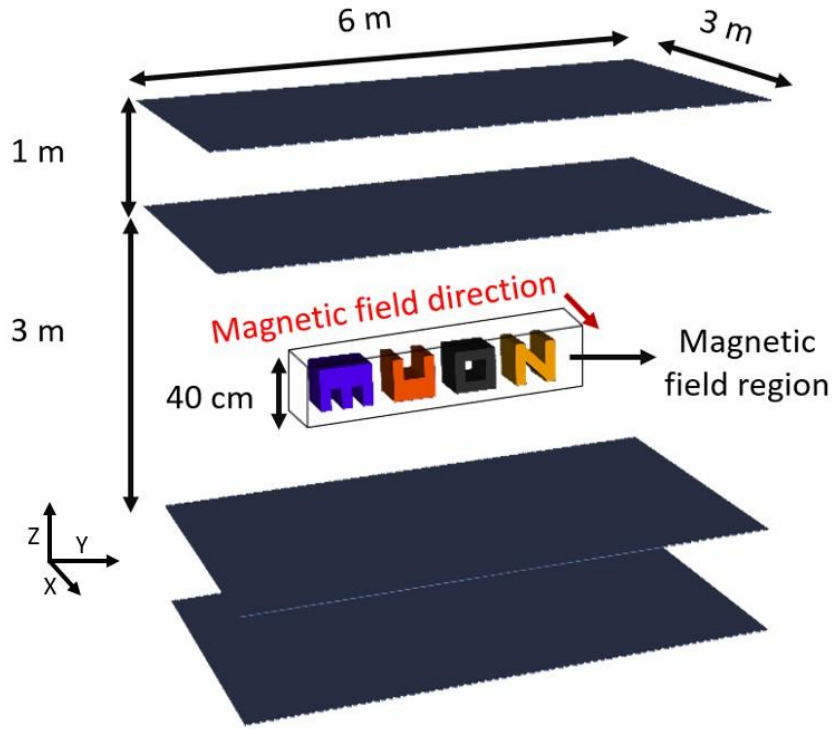

FIGURE 2: Design of the simulation study for the muon scanning system.

## 3. RESULTS

This section presents the findings from our comprehensive simulations and analyses designed to understand the effects of magnetic fields on muon trajectories. We conducted a series of analyses, including the simulation of scattering muography images in the *z-y* plane with integration over the *x*-axis, analysis of scattering angle distributions, examination of the mean scattering angle and coefficient of variation as functions of magnetic field strength, signal-to-noise ratio analysis, and voxelized image analysis under various magnetic field intensities. Through these diverse analytical approaches, we describe how magnetic fields ranging from 0 to 500 mT influence muon scattering patterns, trajectory deviations, and the voxel counts within the investigated volume.

### 3.1. Simulated Scattering Muography Images in z-y Plane

The results presented in Figure 3 demonstrate the impact of different magnetic field strengths on the simulated muon scattering images of the "MUON" target. These images are plotted in the *z-y* plane, integrating over the full 40 cm thickness of the target material along the *x*-axis. Each point in the images represents where a single muon has scattered. The overall pattern we see in each image is made up of these points, showing how all the muons are scattered together under different magnetic field conditions.

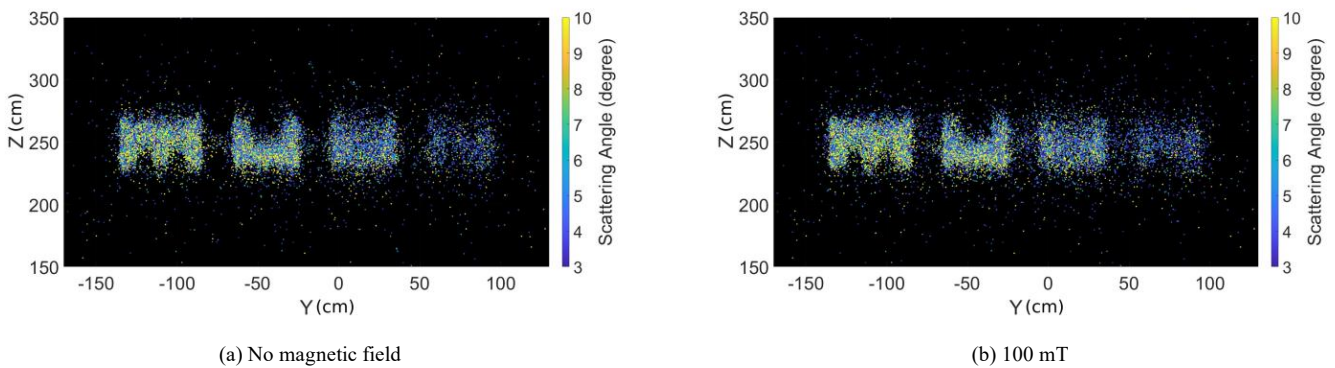

(a) No magnetic field

(b) 100 mT

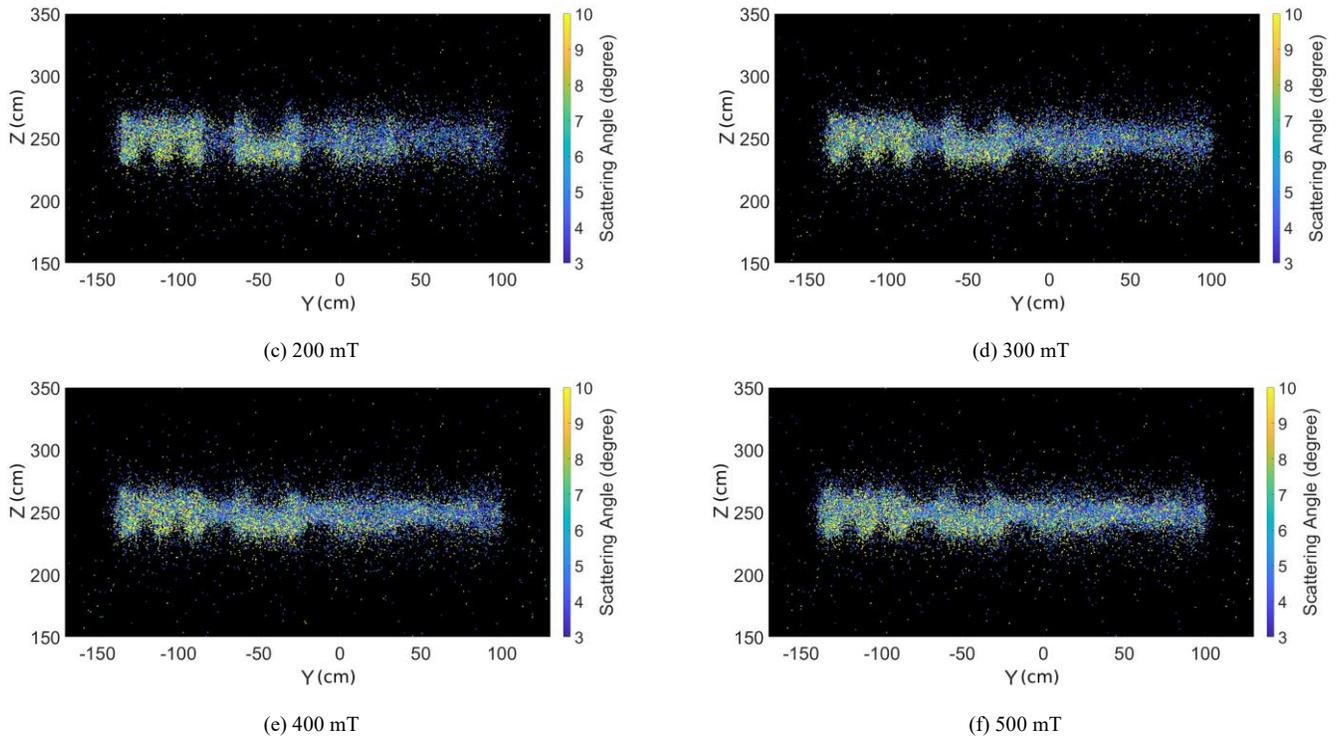

(c) 200 mT  (d) 300 mT

(e) 400 mT  (f) 500 mT

FIGURE 3: Simulated scattering muography images of the "MUON" target under varying magnetic field strengths.

Figure 3(a) shows the simulated scattering muography image of the "MUON" target without any magnetic field, where the clarity of muon scattering is evident. Each pixel's color corresponds to the scattering angle of a single muon after its interaction with the target material.

As the magnetic field is introduced and increased from 100 mT to 500 mT in images (b) through (f), we observe progressive changes in the scattering pattern. The scattering points begin to deviate from the initial distribution seen without a magnetic field, illustrating the influence of the Lorentz force on the muons' trajectories. The images display a more pronounced scattering as the magnetic field strength increases, leading to a more dispersed distribution of scattering points within the $z$-$y$ plane. This distribution is especially noticeable at magnetic field strengths of 300 mT (d), 400 mT (e), and 500 mT (f), where the deviation from the no-field scenario is significant. In the following sections, we will explore our analysis of how magnetic fields influence muon scattering angle.

Figure 4 presents the normalized probability distributions of muon scattering angles under different strengths of magnetic fields. These distributions showcase how the scattering angles of muons change with the application of varying magnetic field intensities. As the magnetic field strength increases, there is a noticeable shift in the distribution of scattering angles, indicating that the magnetic fields are influencing the muon trajectories.

This shift is significant because the scattering angle in our context is not only influenced by the interaction of muons with the target material's atomic nuclei (which depends on the atomic number, $Z$, and the density of the material) but also by the additional deflection caused by the magnetic field. The combined effect results in an increase in the range and probability of larger scattering angles as the magnetic field strength increases.

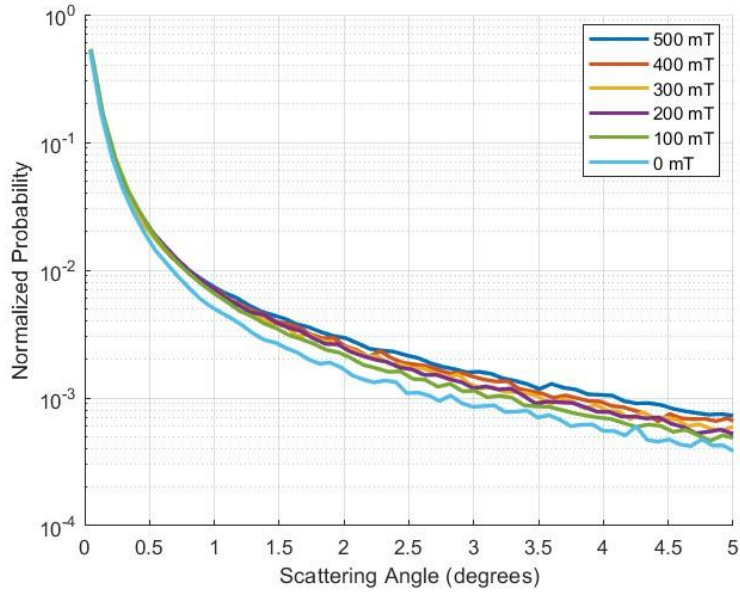

FIGURE 4: Normalized distributions of muon scattering angles under different magnetic field strengths, highlighting the variation in muon deflection behaviors.

*3.2. Mean Scattering Angle vs. Magnetic Field Strength*
The graph in Figure 5 shows how the mean scattering angle of muons changes with different strengths of magnetic fields. It indicates that as the magnetic field strength increases, the average angle at which muons are deflected also increases. This trend demonstrates the effect of magnetic fields on muon trajectories, suggesting that higher magnetic field strengths result in greater muon deflections.

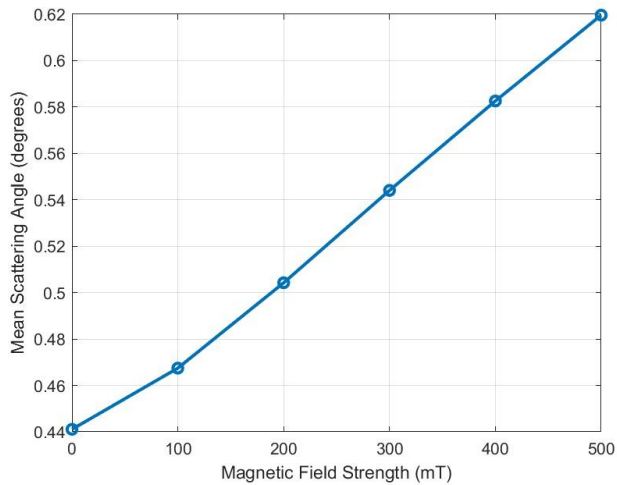

FIGURE 5: Mean scattering angle variation.

*3.3. Coefficient of Variation vs. Magnetic Field Strength*
The coefficient of variation (CV) is calculated as the ratio of the standard deviation ($\sigma$) to the mean ($\mu$) of the scattering angles, expressed as CV = $\sigma/\mu$. This metric offers insights into the scattering angles' spread relative to their average, indicating the degree of variability or uniformity in muon scattering. In our simulation, both scattering and deflection influence the scattering angles of muons. While scattering is a random process, deflection, on the other hand, is a predictable effect caused by magnetic fields that direct muons along more uniform paths in accordance with the Lorentz force.

As the magnetic field strength increases, the influence of deflection on muon trajectories becomes more significant compared to the random scattering effects. This results in a decrease in CV, indicating that the scattering angles become less varied and more uniform as shown in Figure 6. The decrease in CV with stronger magnetic fields suggests that the deflection, being nonrandom and predictable, starts to dominate over the random scattering effects.

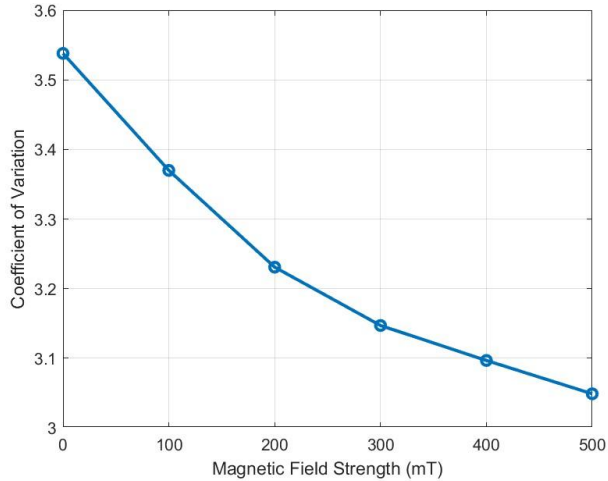

FIGURE 6: Changes in the coefficient of variation with magnetic field strength.

### 3.4. Voxelized Analysis

Voxelization techniques are employed to assess the impact of magnetic fields on muon scattering, where the volume under observation is segmented into cubes of 20 cm × 20 cm × 20 cm. This approach allows for an in-depth study of the changes in muon scattering patterns and their interaction frequencies within each voxel as magnetic field strengths vary.

Figure 7 presents the distribution of muon counts in the x-direction for a specific voxel slice, offering a quantitative view of how the magnetic field affects the accumulation of scattering events within this plane. The data from this figure can elucidate the spatial density of interactions and how it is modified by the presence of varying magnetic field intensities.

In Figure 8, the focus shifts to the mean scattering angles within the same $x$-direction slice, shedding light on the average angular deviation of muons as they are subjected to different magnetic field strengths. This figure is indicative of the magnetic field's effect on the deflection of muons, with the mean scattering angle serving as a proxy for the magnetic field's influence on muon trajectory. Figure 9 illustrates the changes in muon counts for a selected $y$-direction slice, highlighting how the magnetic field distribution influences the scattering across another dimension. This visualization can help in understanding the anisotropy in the interaction pattern, potentially revealing how the field orientation affects muon scattering. Lastly, Figure 10 reveals the muon counts for a slice in the $z$-direction, allowing us to infer how deeply muons penetrate and scatter within the volume and how this scattering is impacted by magnetic fields. The comparison of muon counts at different field strengths can indicate the depth-dependent magnetic effects on muon paths. Together, these figures enable a comprehensive three-dimensional analysis of the magnetic field's effect on muon scattering.

## 4. DISCUSSION

The exploration into the interactions between magnetic fields and muon scattering has unveiled several critical insights, central to advancing the field of muography. Our findings, particularly concerning the phenomenon of "magnetic jamming," highlight the nuanced ways in which magnetic fields can influence muon trajectories and scattering patterns. This influence significantly impacts the clarity, accuracy, and reliability of muography imaging, particularly under varying magnetic conditions.

The effect of magnetic fields on muon detection rates and scattering angles not only complicates material identification and characterization but also challenges the current operational paradigms of muography techniques. As magnetic field strengths increase, the potential for inaccuracies grows, manifesting in the less distinct and identifiable targets within muography images. This is particularly problematic in applications requiring precise material differentiation, such as in geological explorations or security screenings.

Moreover, the implications of magnetic field-induced distortions extend beyond mere technical challenges. They underscore the need for a more comprehensive understanding of muon behavior under the influence of these fields, paving the way for the development of corrective algorithms or compensatory strategies. These strategies are essential for mitigating the effects of magnetic fields, ensuring that muography remains a reliable and precise tool across a broad spectrum of applications, from border security to the identification of nuclear materials.

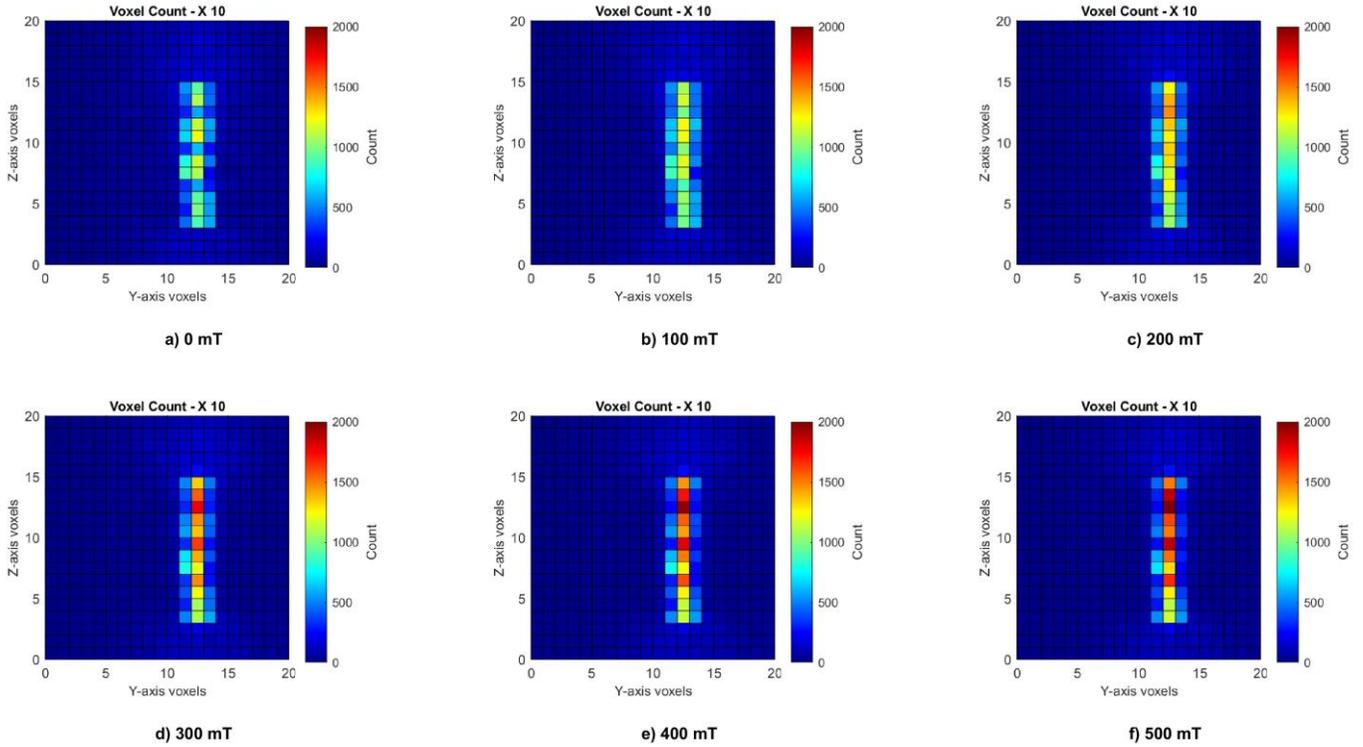

FIGURE 7: Magnetic field impact on the muon counts for slice 10 in the *x*-direction.

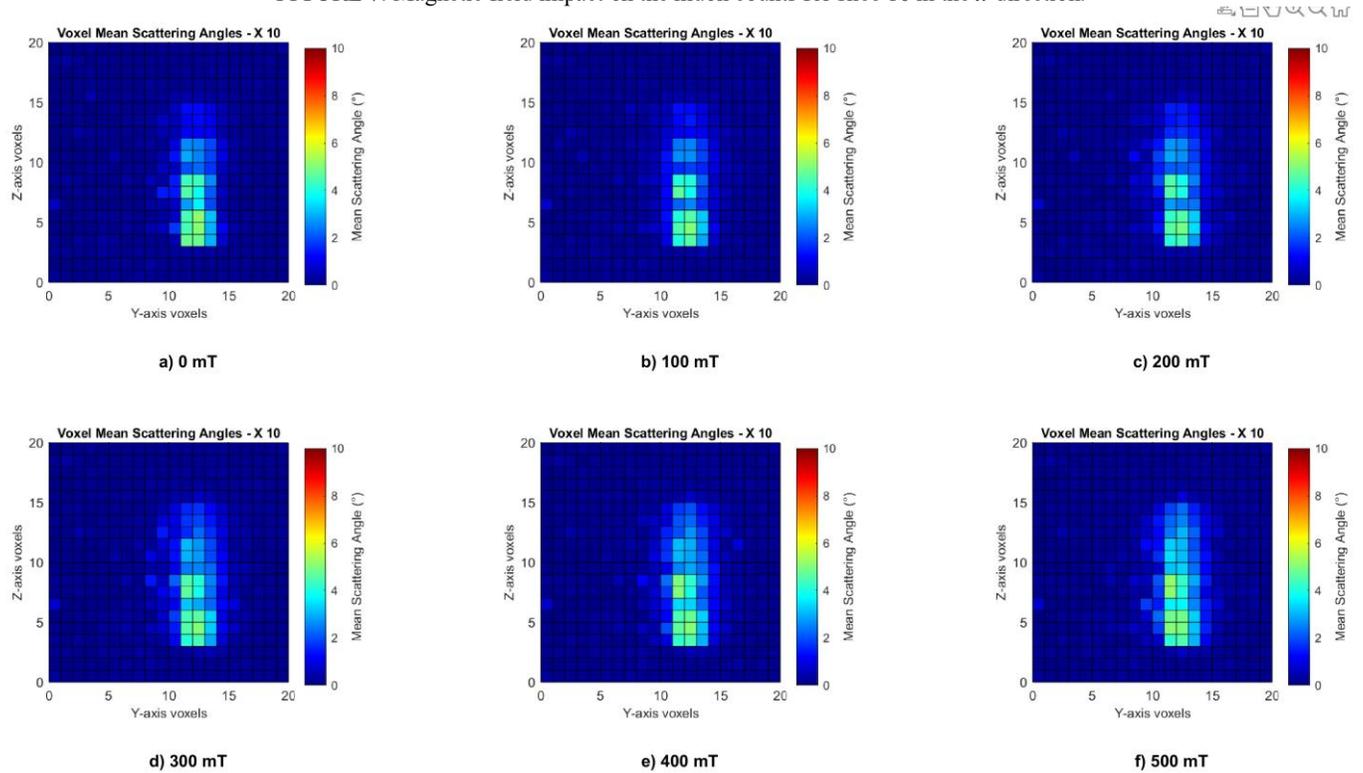

FIGURE 8: Magnetic field impact on the mean scattering angles for slice 10 in the *x*-direction.

## 5. CONCLUSION

Our study has critically examined the relationship between magnetic fields and the performance of muography, employing Monte Carlo simulations to illuminate the impact of these fields on muon path reconstructions and material property inferences. The findings confirm that even moderate magnetic fields can significantly distort muography images, introducing challenges to the accurate interpretation of these images in practical scenarios.

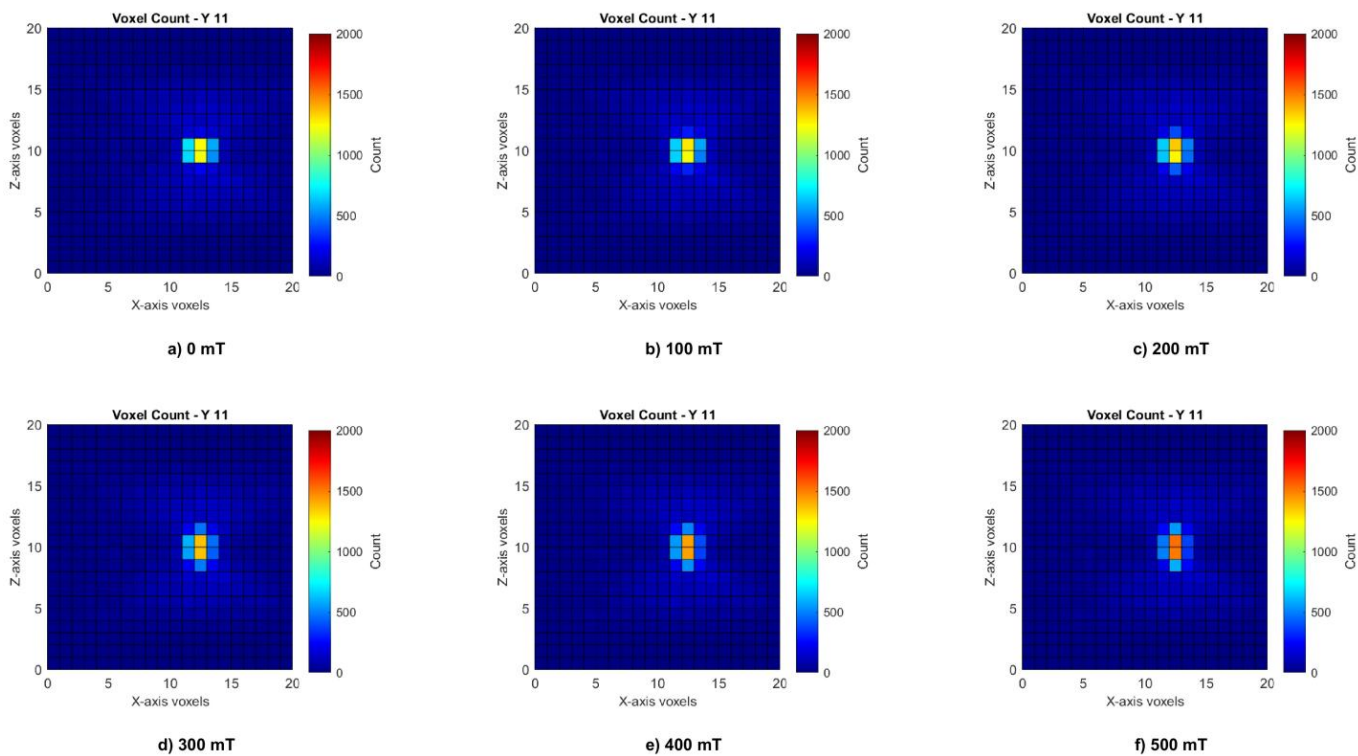

FIGURE 9: Magnetic field impact on the muon counts for slice 11 in the *y*-direction.

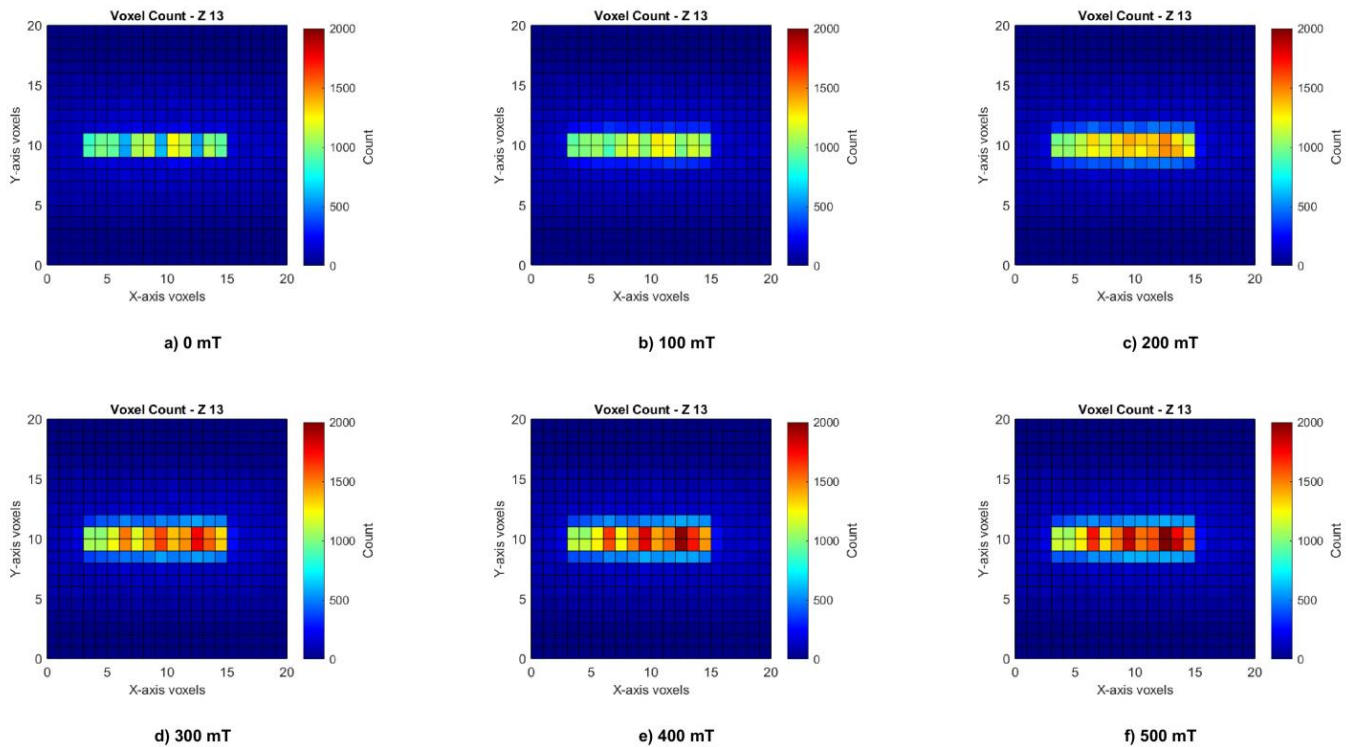

FIGURE 10: Magnetic field impact on the muon counts for slice 13 in the *z*-direction.

This "magnetic jamming" effect, characterized by the distortion of PoCA points and scattering angles, necessitates the development of correction algorithms to adjust for the influence of magnetic fields. The observed increase in scattering angles and the predictability of muon deflections under varying magnetic strengths offer a promising direction for enhancing the accuracy and utility of muography, especially in environments where magnetic fields are prevalent.

Moving forward, the imperative lies in creating mathematical models or computational tools capable of accurately predicting and correcting magnetic distortions. Such advancements will enable muography to overcome the current limitations imposed by magnetic fields, broadening its application and improving its reliability in a range of critical contexts. Through this study, we advocate for continued innovation in muography, aiming to refine its techniques and extend its applicability far into the future.

Although very few "muon portals" are currently operating in the world [3, 20], a hypothetical scenario of intentional "magnetic jamming" is not inconceivable. It is never too late to anticipate the moves of malicious actors and embed the solutions in the design since the onset of a new technology.

## CONFLICTS OF INTEREST

The authors declare that there are no conflicts of interest regarding the publication of this paper.

## ACKNOWLEDGMENTS

This work was partially supported by the EU Horizon 2020 Research and Innovation Programme under grant agreement No. 101021812 ("SilentBorder") and by the Fonds de la Recherche Scientifique (FNRS) under Grants No. T.0099.19 and J.0070.21. We would like to thank Maxime Lagrange for his contributions to this work through weekly meetings and discussions.